\documentclass[aps,prl,preprint,superscriptaddress,showpacs]{revtex4}
\usepackage{graphicx}
\begin{document}
\draft
\def\ee{\varepsilon}
\title{New Type of Fano Resonant Tunneling
via Anderson Impurities in Superlattice}
\author{
S. J. Xu }
\email{sjxu@hkucc.hku.hk}
\affiliation{Department of Physics and HKU-CAS Joint Laboratory on New
  Materials, The University of Hong Kong, Pokfulam Road, Hong Kong,
  China}
\author{S.-J. Xiong}
\affiliation{Department of Physics and HKU-CAS Joint Laboratory on New
  Materials, The University of Hong Kong, Pokfulam Road, Hong Kong,
  China}
\affiliation{National Laboratory of Solid State Microstructures
and Department of Physics, Nanjing University, Nanjing 210093,
China}
\author{J. Liu}
\affiliation{State Key Laboratory for Superlattices and
Microstructures, Institute of Semiconductors, Chinese Academy of
Sciences, Beijing 100083, China}
\author{H. Z. Zheng}
\affiliation{State Key Laboratory for Superlattices and
Microstructures, Institute of Semiconductors, Chinese Academy of
Sciences, Beijing 100083, China}

\author{F. C. Zhang}
\thanks{Leave of absence from Department of Physics, University of Cincinnati, USA.}
\affiliation{Department of Physics and HKU-CAS Joint Laboratory on New
  Materials, The University of Hong Kong, Pokfulam Road, Hong Kong,
  China}


\begin{abstract}

The spectrum of differential tunneling conductance in Si-doped
GaAs/AlAs superlattice is measured at low electric
fields. The conductance spectra feature a zero-bias peak and a low-bias
dip at low temperatures. By taking into account the quantum interference
between tunneling paths via superlattice miniband and via Coulomb
blockade levels of impurities, we theoretically show that such a
peak-dip structure is attributed to a Fano resonance where
the peak always appears at the zero bias and the line shape
is essentially described by a new function $\frac{|\tilde{\varepsilon}|}{|\tilde{\varepsilon}|+1}$
with the asymmetry parameter $q\approx0$. As the temperature increases, the peak-dip
structure fades out due to thermal fluctuations. Good agreement between experiment and
theory enables us to distinguish the zero-bias resonance from the usual Kondo resonance.

\end{abstract}
\maketitle

The Fano effect \cite{c1,fano} is a consequence of interference between
a localized state and a continuum that results in an asymmetric
line shape of the resonance created by this localized state. The
necessary conditions for the observation of the Fano line shape in
a resonance are the coexistence of the localized state and the
continuum, and the quantum coherence between two transition paths.
In condensed-matter physics, the Fano effect was observed in a
wide variety of spectroscopy, such as atomic photoionization,
\cite{c2} optical absorption,\cite{c3} Raman scattering,\cite{c4}
scanning tunneling through a surface impurity atom,\cite{c5,c6}
and phonon-assisted photoluminescence.\cite{xu} Owing to the progress in fabrications of
nanodevices, the system sizes become comparable with the coherence
length and hence the Fano resonance has also been observed in the
transport through mesoscopic systems containing quantum dots
\cite{c7,marcus,c8} and carbon nanotubes.\cite{c9,xie} Moreover, it is
proposed that the Fano resonance can be used as a probe of the
phase coherence in the transport of electrons or other
quasiparticles.\cite{c10,c11} On the other hand, electric-field induced Stark localization,\cite{bastard,mendez} negative differential conductance,\cite{sibble} and Fano resonance due to quantum interference between Stark levels\cite{leo} have been reported in semiconductor superlattices with wide minibands.
In this Letter, we report an observation of
new type of Fano resonance in perpendicular tunneling of electrons through Si-doped
GaAs/AlAs superlattice with 30 periods. The spectrum of
differential conductance at low electric fields shows a clear zero-bias
resonance and a low-bias dip at low temperatures. Simply increasing
the temperature can lead to weakening of such a peak-dip structure, revealing
the coherence nature of the resonance. From the energy structure of the
system, there are two tunneling paths for electrons: the many-body states of Si impurities and the superlattice lowest miniband. Although the hydrogen-like donor
levels are below the bottom of the miniband and have no
contribution to the Fano resonance, the double-occupation levels
of these states are above the Fermi energy due to
the Coulomb blockade effect and can serve as the localized states in
the Fano effect. Starting from this picture we are able to
analytically derive a new line shape function to describe the
observed peak-dip feature where the peak is proved to be always at the zero bias
independent of parameter values and to spread out with increasing temperature.
We also discuss the difference of this zero-bias peak from the usual Kondo resonance.

The sample used in the experiment is a GaAs/AlAs superlattice grown by
molecular beam epitaxy on a $n^+$-type GaAs substrate. The
structure consists of 30 periods of 14 nm GaAs well and 2.5 nm
AlAs barrier. The central 10 nm region of each well is Si doped (4
$\times$ 10$^{17}$ cm$^{-3}$). The key of the designing is in the
precise adjustment of the Si doping density and Al concentration
of the bottom and top AlGaAs contact layers to eliminate the
possible built-in electric field. We minimize the misalignment between the Fermi levels in
the doped superlattice and both contact layers in the absence of the bias.
Standard semiconductor fabrication processes were employed to
fabricate the sample into the $n^+$-$n$-$n^+$ mesa diodes with an
area of 200 $\mu$m $\times$ 200 $\mu$m.

\begin{figure}[h]
\includegraphics[width=13cm]{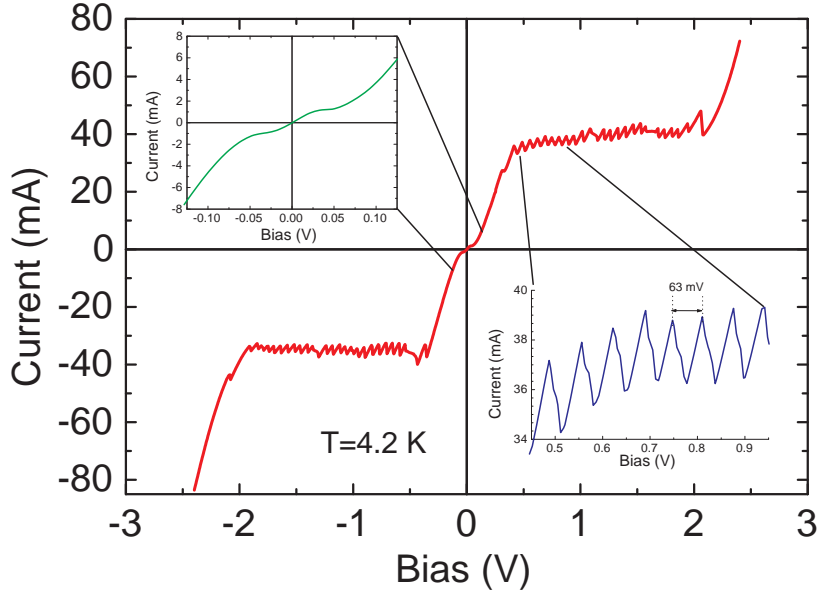}
\caption {(Color online) The current-voltage characteristic of the structured diode
measured at 4.2 K. The inset figures give the enlarged fine structures
in low-field (upper-left inset) and high-field (lower-right inset) regions.}
\label{fig1}
\end{figure}

Figure 1 shows the current-voltage curves of the sample measured
at 4.2 K. Two distinct transport characteristics, as shown in the
inset figures, can be identified. In the high-field region, the
saw-like current oscillations with a period of $\sim$ 60 mV appear
due to the well-known sequential resonance tunneling, also called
high-field domain effect.\cite{13,14} The characteristic in the
low-field region is our main interest in the present work.
Firstly, the linear $I$-$V$ relation within a bias region from
about -10 mV to about 10 mV indicates a finite differential
conductance ($dI/dV$) at zero bias. Secondly, the current exhibits
a saturation in a range about $\pm$40 mV bias, corresponding to a
dip in $dI/dV$ at an average potential drop of 1.3 meV across a
barrier-well supercell. The semilogarithmic differential conductance of the sample
as a function of the bias voltage is plotted in Fig. 2 for various
temperatures. A peak near the zero bias is clearly seen at low
temperatures. In accordance to the current saturation shown in Fig. 1,
the differential conductance exhibits two minima (dips) at the bias voltage
of $\pm 40$ mV. In spite of slight asymmetry probably due to small difference in Fermi level
between the bottom and top contact layers, both dips, especially the dip at +40 mV, are close to zero at 4 K. Furthermore, the dips gradually disappear as temperature increases beyond 100 K.
As mentioned earlier, in each well of the superlattice there exist hydrogen-like orbits at Si
donors and a continuum of two-dimensional (2D) electron states.
Although the levels of hydrogen-like orbits themselves are below
the miniband bottom and have no contribution to the inter-well
tunneling, the levels of double occupation on them should go up due to the Coulomb blockade (CB) effect and are near the
Fermi level. Thus, there are two types of inter-well tunneling
paths, one via the CB levels of Si donors and the other via the superlattice miniband.
Naturally, we can attribute the zero bias peak to the constructive Fano interference between paths
via the CB levels and via the miniband. At the same time, the dips in
differential conductance are originated from the destructive Fano
interference. In Fig. 3 we display the temperature dependence of
the zero-bias differential conductance which shows an interesting temperature dependence. We also depicts the difference in the differential conductances between the peak and the dip
as a function of temperature. The difference drops with increasing temperature as a result of thermal fluctuations. The inset in Fig. 3
shows the line shape fitting curve using the new function derived later. The experimental spectrum can be fitted very well.

\begin{figure}[h]
\includegraphics[width=13cm]{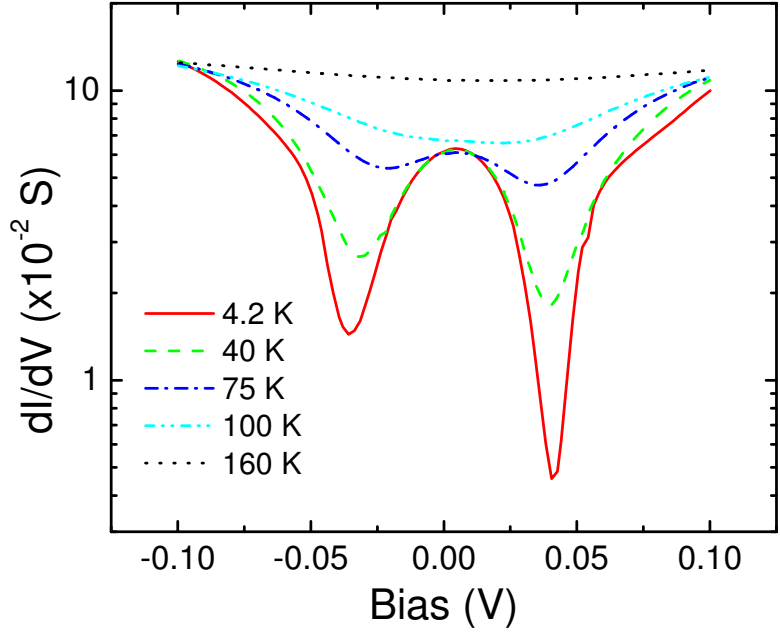}
\caption {(Color online) The measured differential conductance-voltage curves of the
structured diode at different temperatures. The peak at the zero bias and the dip at low bias can be clearly observed at low temperatures.}
\label{fig2}
\end{figure}

\begin{figure}[h]
\includegraphics[width=13cm]{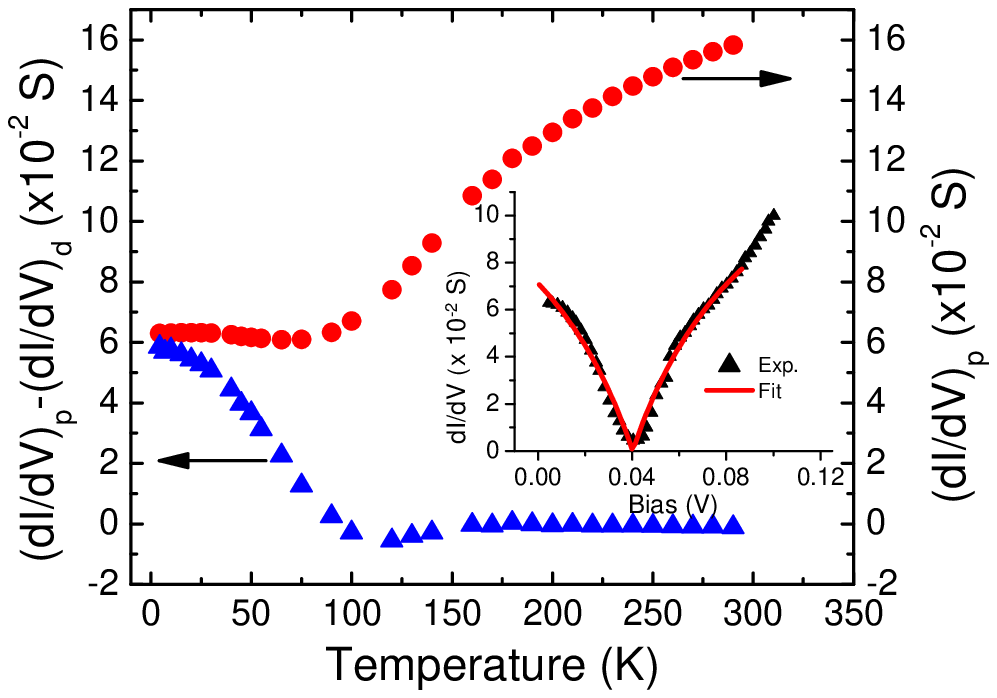}
\caption {(Color online) The measured differential conductance dI/dV at zero bias (solid circles) as a
function of temperature. The temperature dependence of the
difference in dI/dV between the peak and the dip (solid triangles) is also depicted. The inset shows
the line shape fitting curve (solid line) to the measured spectrum (solid triangles) with the newly derived function $\frac{|\tilde{\varepsilon}|}{|\tilde{\varepsilon}|+1}$
where $\tilde{\varepsilon}=(V_b-a)/b$ with $V_b$ being the bias, $a=(0.0402\pm 0.0003)$V, and $b=(0.059\pm 0.008)$V.
The total variance of fitting is $\chi^2 =10^{-5}$.} \label{fig3}
\end{figure}

The basic features of low-bias behavior shown in Figs. 1 and 2
are: (1) At low temperatures there is a peak at the zero bias and
a cuspidal dip at a finite bias in curves of differential
conductance versus bias voltage; (2) The peak-dip structure is
gradually smoothed by increasing the temperature. Theoretically,
the transport in such systems could be attributed to impurity
assisted tunneling or to Kondo resonant tunneling. However, the
mechanism of impurity assisted tunneling usually gives peaks at
finite bias and can not explain the dip structure.\cite{aaa} On
the other hand, although the Kondo effect can provide a peak at
the zero bias, the peak width and the temperature are too snall,
inconsistent with the observed results. Furthermore, in the system
many impurities are involved, this makes the zero-bias peak become
much narrower.\cite{bbb} We now turn to a new theoretical analysis
for this observed peak-dip structure. The motion of electrons in
the perpendicular direction of the superlattice at low bias is
related to the minibinds which are formed by the coupling between
2D electron continuum in adjacent quantum wells. The energy of an
electron within a quantum well can be written as
\begin{equation}
   E_{i,{\bf k}_{\parallel}} = 2 t_0 (\cos k_x +\cos k_y) +K_i,
   \end{equation}
where ${\bf k}_{\parallel}\equiv (k_x,k_y) $ is the dimensionless
wave vector in the parallel plane, $K_i$ is the quantized
energy in the perpendicular direction with $i$ being the level
index, and $t_0$ is the hopping energy describing the in-plane
motion in the tight-binding representation. $t_0$ can be estimated
by $t_0 \sim \frac{\hbar^2}{2m^*a_0^2}$ with $m^*$ being the
effective mass and $a_0$ the lattice spacing. In the wells $m^*
\sim 0.067 m_0$,\cite{18} with $m_0$ the free electron mass. On
the other hand, the Si atoms doped in the middle region of each well become
hydrogen-like centers with positive charge after
contributing electrons. Because the wells are wide enough ($\sim
140$ \AA), we can neglect the effect of the well structure on the
hydrogen-like states, which are bound to silicon impurities with a
binding energy $\epsilon_0 \sim$ -7 meV and a radius of $r_0$ $\sim$
88 $\text{\AA}$ estimated from the effective mass and a dielectric
constant $\epsilon_s =11$. The Coulomb repulsion energy for double
occupation is estimated from the size to be $U \sim
\frac{e^2}{\epsilon_s r_0} \sim 14$ meV. Thus, for a quantum well
one has a schematic energy spectrum illustrated by the density of
states in Fig. 4. The electrons are donated by Si impurities in the
wells and in the contact layers. They fill up the hydrogen-like
orbits and the lowest 2D continuum to form a Fermi level $E_f$.
Thus, $E_f$ is usually between the bottom of the 2D continuum and
the double occupation level, but its precise position is
influenced by other physical ingredients such as the bend of the
potential and the doping in contact layers.

\begin{figure}[h]
\includegraphics[width=13cm]{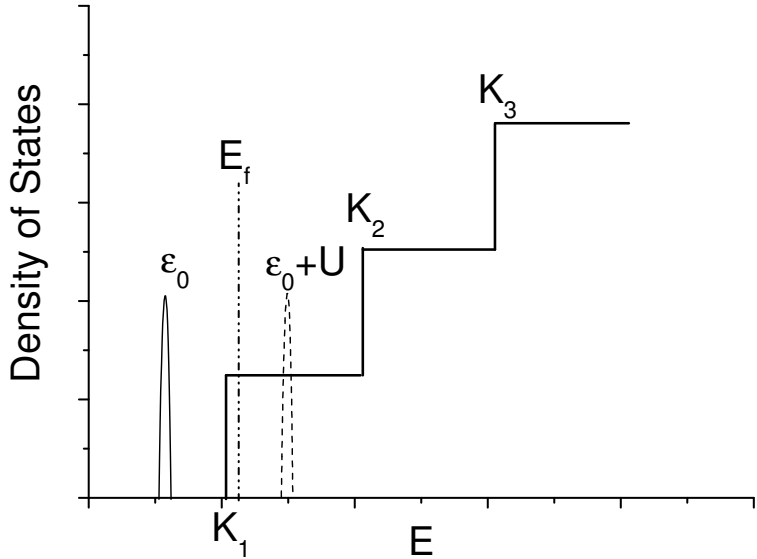}
\caption {Schematic illustration of density of states in a quantum
well with doped Anderson impurities. The Fermi level $E_f$ of the system, the hydrogen-like state $\epsilon_0$ and Coulomb blockade state $\epsilon_0+U$ of impurities are illustrated.} \label{fig4}
\end{figure}

By introducing the inter-well tunneling in superlattice, the quantized
energy in isolated wells is extended to a miniband as one of the major transport paths in
the perpendicular direction. Due to the existence of the Si
impurities and inevitable roughness at interfaces, the parallel
momenta are not necessarily conserved in the inter-well tunneling.
On the other hand, the widths of wells and barriers in the superlattice
make the lowest superlattice miniband in the perpendicular
direction certainly narrow (only 0.3 meV as
calculated from the Kronig-Penny model of superlattice potential profile). Thus, at low
temperatures and low but finite bias (larger than the order of
meV), the miniband tunneling is based on the resonance in the
perpendicular direction but without strict matching of parallel
momenta between the adjacent wells. In fact, when an electron is tunneling from
one well to a sequent well due to a bias between them, the
kinetic energy in the perpendicular direction should be kept
almost the same so that the sharp resonance maintains, and the
energy gained from the bias should be transferred to the
kinetic energy in the parallel direction. This kind of energy transfer
can take place due to elastic scattering from the interface roughness
or impurities which makes inter-well tunneling between different
parallel momenta possible. On the contrary, the contribution from
the completely coherent tunneling with conservation of parallel
momenta is vanishingly small owing to the narrow miniband width in
the perpendicular direction smaller than the inter-well bias.
Since the parallel momenta are not fixed, the tunneling spectrum
is determined by the density of states of wells near the Fermi
level, rather than by the miniband structure in the perpendicular
direction. As can be seen from Fig. 4, the spectral density in a
well near the Fermi surface is certainly flat, giving only
structure-less tunneling spectrum. So the observed peak-dip
structure in differential conductance at low bias is a more distinct
effect, suggesting a specific interfering mechanism. Actually, besides
the miniband transport discussed above, there is another path for perpendicular motion of
electrons, i.e., the path via the hydrogen-like states of Si atoms. In
the present case the latter is related to the Coulomb blockade
levels of these states.

Below we use a tight-binding chain with sites at centers of
barriers to describe the
perpendicular motion in the miniband. The site spacing is defined as the size of a supercell,
and the hopping $t_1$ between adjacent sites in the chain
is associated with the miniband width. In this tight-binding representation, model Hamiltonian for
the perpendicular motion through a supercell can be written as:
\begin{equation}
    H_T = \sum_{{\bf k}_\parallel,{\bf k}'_\parallel} t_1 (a^\dag_{{\bf k}_\parallel,1} a_{{\bf k}'_\parallel,2}+
    \text{H.c.})
    +\sum_{n,{\bf k}_\parallel}
    (t'_{1 ;n} a_{{\bf k}_\parallel,1}^\dag d_n +t'_{2 ;n} a_{{\bf k}_\parallel,2}^\dag d_n
    + \text{H.c.}),
\end{equation}
where $({\bf k}_\parallel,1)$ and $({\bf k}'_\parallel,2)$ denote
the states in two ends of the supercell, $a^\dag_{{\bf
k}_\parallel,1(2)}$ and $d^\dag_n$ are creation operators of
electrons in state $({\bf k}_\parallel,1(2))$ and in the $n$th
hydrogen-like orbit, respectively, and $t'_{i;n}$ is the hopping
between state at the $i$th end of the well and the hydrogen-like
orbit $n$. Note that in the Hamiltonian the strict conservation of
the parallel momentum is not necessary for the perpendicular
resonant transport. As mentioned above, the resonant tunneling
within the narrow miniband in the perpendicular direction is kept.
Thus, when an electron is injected from one end of the supercell
under a bias $\Delta$, its wave function can be expressed by a
superposition
\begin{equation}
   \Psi = \left(\sum_{j\leq 1} (e^{i k z_j} +r e^{-i k z_j})|z_j \rangle
   \right) \Phi(E_\parallel-\Delta)+ \left(\sum_{j\geq 2} t e^{i k z_j}|z_j \rangle
   \right)\Phi(E_\parallel) + \sum_n x_n |d_n\rangle,
   \end{equation}
where $r$ and $t$ are amplitudes of reflected and transmitted
waves, $\Phi(E_\parallel)$ is the parallel part of wave function
with the parallel kinetic energy within a small bin at
$E_\parallel$, $z_j$ is the position of the $j$th site in the
chain, $k$ is the wave vector in the perpendicular direction, and
$x_n$ is the amplitude on impurity $n$. By applying the
Hamiltonian on $\Psi$, we obtain the following equations:
\begin{equation}
   \tilde{t}_1 (e^{\text{i}k} + r e^{-\text{i}k}) =  \tilde{t}_1 t + \sum_n
   \tilde{t'}_{1;n} x_n,
   \label{qq1}
   \end{equation}
\begin{equation}
   E x_n = (\epsilon_0+U) x_n + \tilde{t'}^*_{1;n} (1+r)
   + \tilde{t'}^*_{2;n} t,
   \label{qq2}
   \end{equation}
   \begin{equation}
   \tilde{t}_1 t e^{-\text{i}k}  =  \tilde{t}_1 (1+r) + \sum_n
   \tilde{t'}_{2;n} x_n,
   \label{qq3}
   \end{equation}
where $E$ is the energy of electron, $\tilde{t}_1 = \rho_0 t_1$
and $\tilde{t'}_{1(2);n} = \sqrt{\rho_0} {t'}_{1(2);n}$ with
$\rho_0$ being the number of 2D states in the energy bin. Here, we
omit the spin index as electrons with up and down spins have the
same probability to tunnel through two paths if there is no spin
polarization in impurities. The kinetic energy in parallel
directions does not appear in the equations since it is restricted
by energy $E$ and perpendicular wave vector $k$. Because the
distribution of Si impurities is symmetric about the center of the
well, it is reasonable to assume that the average hopping from Si
impurities to both ends is symmetric, namely, $\sum_n
|{t'}_{1;n}|^2 =\sum_n |{t'}_{2;n}|^2=\sum_n
{t'}^*_{1;n}{t'}_{2;n} \equiv V t_1$. Solving Eqs. (\ref{qq1}),
(\ref{qq2}), and (\ref{qq3}) yields the transmission coefficient
as
 \begin{equation}
  \label{transmi}
 T(E) \equiv   |t|^2 = \frac{( E-\epsilon_0 -U +V)^2 \cos^2
 \frac{k}{2}}{ (E-\epsilon_0-U) (E-\epsilon_0-U+2V) \cos^2
 \frac{k}{2} + V^2} .
  \end{equation}
Since the bias applied to the superlattice is much larger than the
width of superlattice miniband, it can be assumed that the entire
miniband is within the energy window spanned by the bias. So the
inter-well transmission is determined by the whole miniband and
the averaging over all $k$ has to be done. After performing this
average, we have
\begin{equation}
 \label{tt}
  \bar{T}(E) =
  \frac{|E-\epsilon_0-U +V|}{\left|
  E-\epsilon_0-U+V
  \right| +\left|V \right|}.
  \end{equation}

The peak-dip structure in $\bar{T}(E)$ appears when $V\neq 0$,
otherwise it is only a constant. From the transmission coefficient
one can calculate the tunneling current through a supercell with
the Landauer formula
\begin{equation}
 I = \frac{eM}{h} \int dE \bar{T}(E)
 [f(E-\Delta )- f(E)],
 \end{equation}
where $M$ is related to the in-plane area of the system and
\[
 f(E) =\frac{1}{\exp\left( \frac{E-E_f}{k_BT} \right) +1}
 \]
is the Fermi-Dirac statistical distribution. One has the
bias-dependent differential conductance
\begin{equation}
 \label{gg}
g \equiv e\frac{d I}{d\Delta} = -\frac{Me^2}{h} \int dE \bar{T}(E)
 \frac{\partial f(E-\Delta)}{\partial E}.
 \end{equation}

By taking values of parameters $\epsilon_0$, $E_f$, $U$, and $V$
estimated above, we calculate the bias and temperature dependence
of differential conductance and show the results in Fig. 5. It is
interesting to note that the peak-dip structure is well reproduced
at low temperatures. By increasing the temperature, the peak-dip
structure is rapidly removed. The characteristic temperature for
this transition is in the order of several tens of Kelvin. This is
also in agreement with the experiment.

\begin{figure}[h]
\includegraphics[width=13cm]{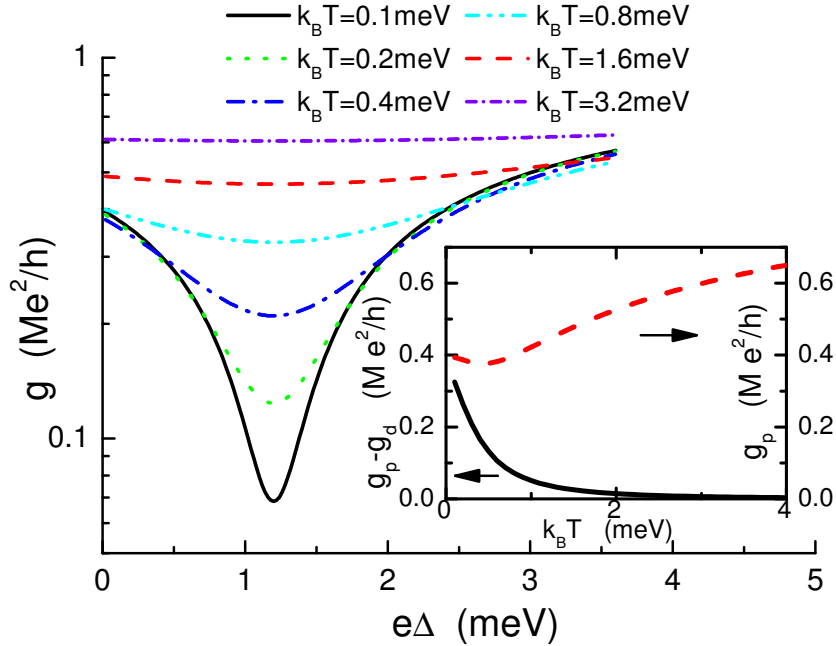}
\caption {(Color online) Calculated differential conductance as a
function of the bias at various temperatures. The parameters are:
$\epsilon_0=-7$ meV, $E_f=4$ meV, $U= 14$ meV, and $V=1.8$ meV. In
the inset the temperature dependence of the peak differential
conductance (dashed line) and the difference in differential conductance between
the peak and the dip (solid line) is plotted.
   } \label{fig5}
\end{figure}

From Eqs. (\ref{tt}) and (\ref{gg}), at low temperatures one has
the following function to describe the asymmetric line shape in
bias dependence of differential conductance
\begin{equation}
\label{qq}
   g(\Delta ) \propto
   \frac{|\tilde{\varepsilon}|}{|\tilde{\varepsilon}| +1},
   \end{equation}
where $\tilde{\varepsilon}$ is the renormalized bias:
\[
 \tilde{\varepsilon} =\frac{ \Delta -\epsilon_0 -U+V}{V}.
 \]

Although the above formulas are derived for one supercell, the
same results can be obtained for other supercells because each of
them has the same drop of the Fermi energy providing the same
energy window for the tunneling electrons. This ensures the same
current in any parts of the chain. It is interesting to note that
the line-shape function of Eq. (\ref{qq}) is different from the
standard Fano line shape function
$\frac{(\tilde{\varepsilon}+q)^2}{\tilde{\varepsilon}^2+1}$.
Especially, from Eq. (\ref{qq}) the peak is always at the zero
bias ($\Delta =0$) as long as $\epsilon_0+U -V > 0$. Although this
zero-bias peak seems like the Kondo resonance, it is much
wider, and is more robust against temperature. So the Kondo
effect could be excluded from the possible mechanisms for the observed zero-bias
peak.

It should be emphasized that the miniband transport here is different from
that in short-period superlattices with much wider miniband, e.g., 30 meV or larger.\cite{sibble}
In the latter the miniband transport is based on the coherent inter-well tunneling with
conserved parallel momentum manifested by the negative differential conductance \cite{sibble} or the Wannier-Stark
localization \cite{mendez} under an electric field. In the present case the miniband is
extremely narrow so that the effective inter-well tunneling originates only from processes without
conservation of parallel momentum. Interfering with the tunneling paths via the Coulomb blockade
levels of Anderson impurities, the transport paths provided by the narrow miniband as a whole
result in the observed special Fano resonance
with line shape described by function of Eq. (\ref{qq}).

In conclusion, we have measured conductance of silicon doped
GaAs/AlAs superlattice along the direction perpendicular
to the layers. The differential conductance shows a zero-bias peak
accompanied with nearby dips at low temperatures. These features
suggest that the Fano interference between miniband transport and
transport via hydrogen-like orbits on Si impurities dominates the
electron transport at low temperatures and at low electric-fields.
The observed peak-dip structure and its temperature dependence can
be well reproduced by calculations of differential conductance on
basis of quantum interference of two transmission channels.

{\it Acknowledgments}: Main experimental work was carried out at
State Key Laboratory for Superlattices and Microstructures,
Institute of Semiconductors, Beijing. We wish to thank H. Guo, T.
K. Ng, and X. C. Xie for valuable discussions. The authors
gratefully acknowledge X. P. Yang and P. H. Zhang for the growth
of the sample, C. F. Li for processing the sample, and Y. X. Li
for her contribution in the measurements of the low-temperature
$I$-$V$ relation and of conductance spectra. S-JX wishes to thank
research support from Chinese National Foundation of Nature
Sciences (Nos. 60276005 and 10474033). H. Z. Zheng wishes to thank
the supports from Major State Basic Research Project (No.
G001CB3095), Special Project from Chinese Academy of Sciences.


\begin{references}

\bibitem{c1} U. Fano, Phys. Rev. {\bf 124}, 1866 (1961).

\bibitem{fano} U. Fano and J. W. Cooper, Phys. Rev. {\bf 137}, A1364 (1965); Rev. Mod. Phys. {\bf40}, 441 (1968).

\bibitem{c2} U. Fano and A. R. P. Rau, {\it Atomic Collision
and Spectra}, (Academic Press, Orlando, 1986).

\bibitem{c3} J. Faist, F. Capasso, C. Sirtori, K. W. West, and L. N.
Pfeiffer, Nature (London) {\bf 390}, 589 (1997).

\bibitem{c4} F. Cerdeira, T. A. Fjeldly, and M. Cardona, Phys. Rev. B
{\bf 8}, 4734 (1973).

\bibitem{c5} V. Madhavan, W. Chen, T. Jamneala, M. F. Crommie, and N. S.
Wingreen, Science {\bf 280}, 567 (1998).

\bibitem{c6} J. Li, W.-D. Schneider, R. Berndt, and B. Delley, Phys. Rev. Lett.
{\bf 80}, 2893 (1998).

\bibitem{xu} S. J. Xu, S.-J. Xiong, and S. L. Shi, J. Chem. Phys. {\bf123},
221105 (2005).

\bibitem{c7} K. Kobayashi, H. Aikawa, S. Katsumoto, and Y. Iye, Phys. Rev.
Lett. {\bf 88}, 256806 (2002).

\bibitem{marcus}A. C. Johnson, C. M. Marcus, M. P. Hanson, and A. C. Gossard, Phys. Rev.
Lett. {\bf 93}, 106803 (2004).

\bibitem{c8} M. Sato, H. Aikawa, K. Kobayashi, S. Katsumoto, and Y. Iye,
Phys. Rev. Lett. {\bf 95}, 066801 (2005).

\bibitem{c9} J. Kim, J.-R. Kim, J.-O. Lee, J. W. Park, H. M. So, N. Kim,
K. Kang, K.-H. Yoo, and J.-J. Kim, Phys. Rev. Lett. {\bf 90},
166403 (2003).

\bibitem{xie} W. Yi, L. Lu, H. Hu, Z. W. Pan, and S. S. Xie, Phys. Rev. Lett. {\bf91}, 076801 (2003).

\bibitem{c10} A. A. Clerk, X. Waintal, and P. W. Brouwer, Phys. Rev. Lett. {\bf 86},
4636 (2001).

\bibitem{c11} Y.-J. Xiong and S.-J. Xiong, Int. J. Mod. Phys. B {\bf 16}, 1479
(2002).

\bibitem{bastard} J. Bleuse, G. Bastard, and P. Voisin, Phys. Rev. Lett. {\bf 60}, 220 (1988).

\bibitem{mendez} E. E. Mendez, F. Agull\'{o}-Rueda, and J. M. Hong, Phys. Rev. Lett. {\bf 60}, 2426 (1988).

\bibitem{sibble} A. Sibille, J. F. Palmier, H. Wang, and F. Mollot, Phys. Rev. Lett. {\bf 64}, 52 (1990).

\bibitem{leo} C. P. Holfeld, F. L\"{o}ser, M. Sudzius, K. Leo, D. M. Whittaker, and K. K\"{o}hler, Phys. Rev. Lett. {\bf 81}, 874 (1998).

\bibitem{13} K. K. Choi, B. F. Levine, R. J. Malik, J. Walker, and C. G. Bethea, Phys. Rev. B {\bf35},
4172 (1987).

\bibitem{14} H. T. Grahn, R. Haug, W. Muller, and K. Ploog, Phys. Rev. Lett. {\bf 67},
1618 (1991).

\bibitem{aaa} S. J. Xu, {\it et al.}, Phys. Lett. A

\bibitem{bbb} Qing-Qiang Xu and Shi-Jie Xiong, unpublished. 

\bibitem{18} S. S. Allen and S. L. Richardson, Phys. Rev. B {\bf 50}, 11693
(1994).

\end{references}
\end{document}